# Теорія та методика навчання математики, фізики, інформатики

*Збірник наукових праць*
*Випуск VIII*

Том 1

Збірник містить статті з різних аспектів дидактики математики і проблем її викладання у ВНЗ та школі. Значну увагу приділено питанням розвитку методичних систем навчання математики та модернізації математичної освіти в контексті орієнтирів Болонського процесу.

Для студентів вищих навчальних закладів, аспірантів, наукових та педагогічних працівників.

Редакційна колегія:

*В.М. Соловйов*, доктор фізико-математичних наук, професор
*М.І. Жалдак*, доктор педагогічних наук, професор, акад. АПН України
*Ю.С. Рамський*, кандидат фізико-математичних наук, професор
*В.І. Клочко*, доктор педагогічних наук, професор
*С.А. Раков*, доктор педагогічних наук, професор
*Ю.В. Триус*, доктор педагогічних наук, професор
*П.С. Атаманчук*, доктор педагогічних наук, професор
*В.Ю. Биков*, доктор технічних наук, професор, чл.-кор. АПН України
*О.Д. Учитель*, доктор технічних наук, професор
*І.О. Теплицький*, кандидат педагогічних наук, доцент (відповідальний редактор)
*С.О. Семеріков*, кандидат педагогічних наук, доцент (відповідальний редактор)

Рецензенти:

*Г.Ю. Маклаков* – д-р техн. наук, професор кафедри інформаційних технологій навчання Севастопольського міського гуманітарного університету, науковий керівник лабораторії розподілених систем навчання та дистанційної освіти

*А.Ю. Ків* – д-р фіз.-мат. наук, професор, завідувач кафедри фізичного та математичного моделювання Південноукраїнського національного педагогічного університету ім. К.Д. Ушинського (м. Одеса)

# ОСОБЛИВОСТІ ЗАСТОСУВАННЯ ММС SAGE В МОБІЛЬНОМУ КУРСІ ВИЩОЇ МАТЕМАТИКИ

К.І. Словак[1], М.В. Попель[2]
[1] м. Кривий Ріг, Криворізький економічний інститут Київського національного економічного університету імені Вадима Гетьмана
[2] м. Кривий Ріг, Криворізький державний педагогічний університет
Slovak_kat@mail.ru

Сучасне уявлення про якісну математичну освіту передбачає таку важливу складову навчального процесу, як застосування інформаційно-комунікаційних технологій математичного призначення, зокрема – систем комп’ютерної математики (СКМ). Використання СКМ у курсі вищої математики дозволяє: більш наочно і зрозуміло подати теоретичний матеріал; позбавити студентів від виконання рутинних обчислень; забезпечити багаторівневий процес навчання, а, отже, сприяє підвищенню пізнавального інтересу і головне – дозволяє зробити процес навчання більш швидким та змістовним.

Поступово все більшої популярності набуває новий напрямок розвитку СКМ – мобільні математичні середовища (ММС).

*Мобільне математичне середовище* – це мережне програмне забезпечення, що надає можливість доступу до математичних об’єктів в будь-який зручний час та будь-який спосіб. Застосування таких середовищ дозволяє інтегрувати аудиторну і позааудиторну роботу у безперервний навчальний процес та надає можливість організувати в межах одного середовища повний цикл навчання: а) зберігання та подання навчальних матеріалів; б) математичних досліджень; в) індивідуальної та колективної роботи; г) оцінювання навчальних досягнень.

Яскравим представником мобільних математичних систем є Web-СКМ Sage.

За допомогою Sage можна:

1) виконувати будь-які обчислення, як аналітичні (дії з алгебраїчними виразами, розв’язування рівнянь, диференціювання, інтегрування тощо), так і чисельні (точні – з будь-якою розрядністю, наближені – з будь-якою, наперед заданою точністю);

2) подавати результати обчислень у зручній для сприйняття формі, будувати дво- та тривимірні графіки кривих та поверхонь, гістограми та будь-які інші зображення (в тому числі анімаційні);

3) поєднувати обчислення, текст та графіку на робочих листах з можливістю їх друку, оприлюднення в мережі та спільної роботи над ними;

4) створювати за допомогою вбудованої у Sage мови Python моделі для виконання навчальних досліджень;

5) створювати нові функції та класи мовою Python [1].

Під час вивчення курсу вищої математики ММС Sage доцільно використовувати за такими напрямами:

– графічні інтерпретації математичних моделей та теоретичних понять;
– автоматизація рутинних обчислень;
– підтримка самостійної роботи;
– математичні дослідження [2].

Таким чином, ММС Sage можна вважати ефективним засобом для створення мобільних курсів, а також модульним динамічним об'єктно-орієнтованим середовищем для навчання.

В процесі вивчення мобільного курсу вищої математики особливої уваги заслуговує застосування моделей з графічним інтерфейсом і напівавтоматичним управлінням. Використання та дослідження таких моделей дозволяє значно легше зрозуміти математичну, фізичну чи економічну суть методів та алгоритмів; глибше усвідомити новий матеріал та створити змістову основу для розв'язання прикладних задач, а також сприяє підвищенню пізнавальної активності через наочність.

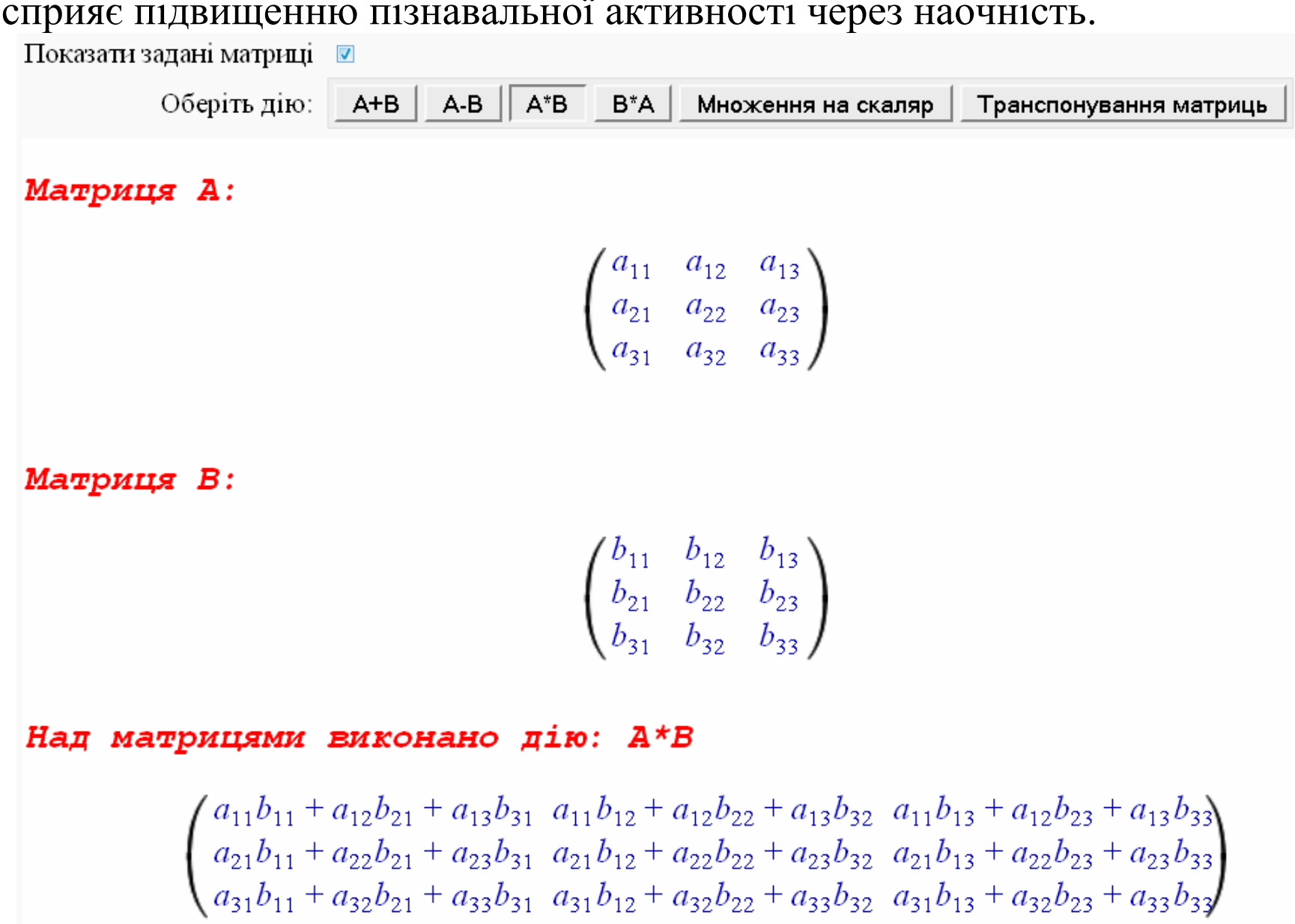

Рис. 1. Інтерфейс користувача моделі «Операції над матрицями»

Наведемо декілька прикладів.

Так, під час вивчення першого модуля курсу «Елементи лінійної алгебри», зокрема теми «Матриці та дії над ними», пропонуємо студентам модель, яка демонструє правила додавання та віднімання матриць, множення матриці на скаляр, транспонування матриці, множення двох матриць на прикладі квадратних матриць третього порядку (рис. 1).

Особливістю цієї моделі є те, що результат виконання тієї чи іншої операції подано у вигляді формули, тобто користувач не просто отримує готовий результат, а бачить, які дії потрібно виконати для того, щоб отримати суму, різницю чи добуток матриць.

Таку модель доцільно використати під час лекції, звільняючи викладача від громіздких записів на дошці, а студентів у зошитах, тим самим вивільняючи час на обміркування та засвоєння алгоритмів розв’язування задач.

При вивченні модуля «Елементи векторної алгебри» пропонуємо застосовувати дві моделі: перша – ілюструє операції над векторами (рис. 2), друга – залежність скалярного добутку векторів від градусної міри кута між ними (рис. 3). Під час роботи з останньою моделлю змінюючи кут між векторами (рухаючи повзунок відповідного поля) студент переконується, що дійсно скалярний добуток перпендикулярних векторів дорівнює нулю тощо.

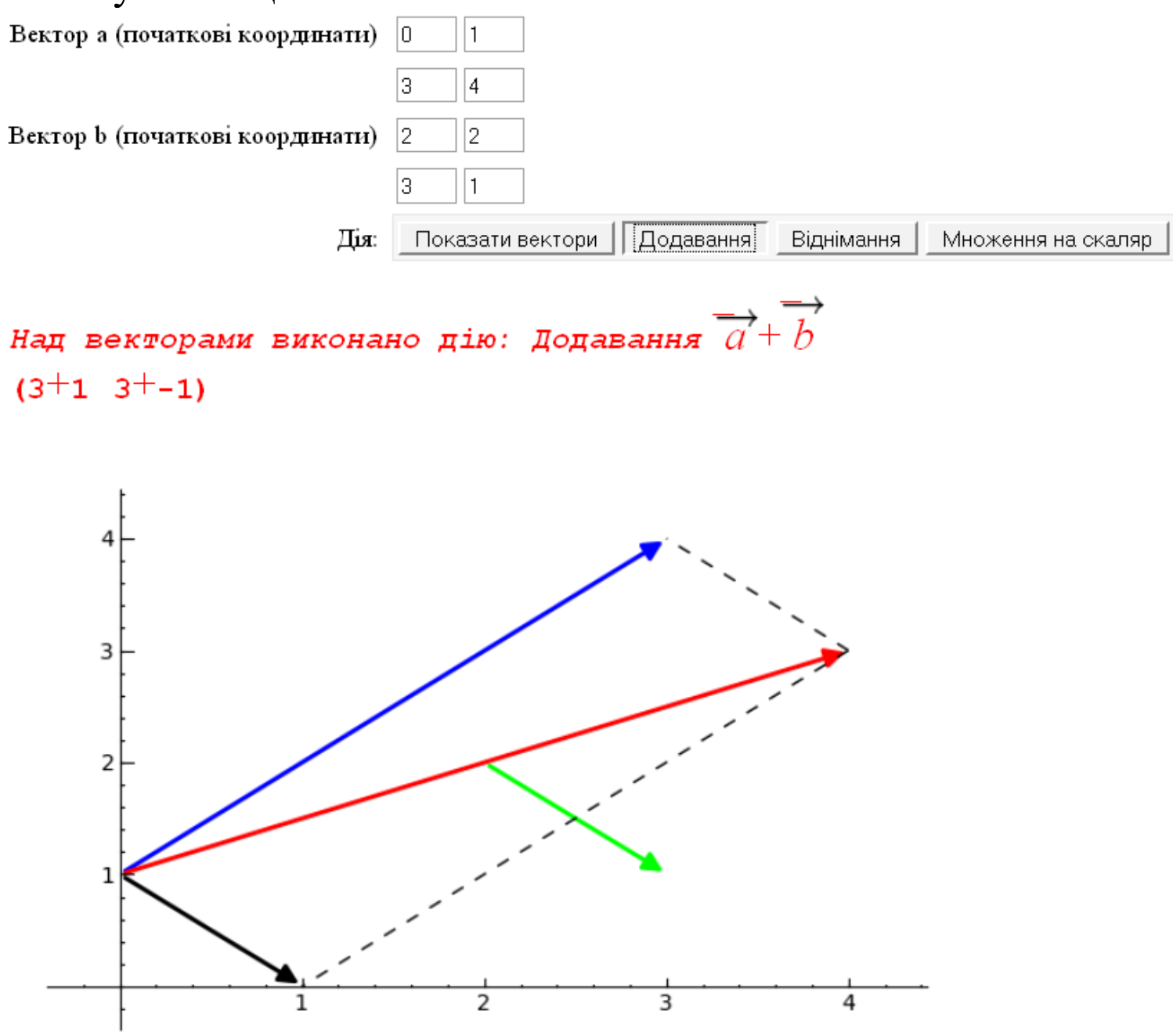

Рис. 2. Інтерфейс користувача моделі «Операції над векторами»

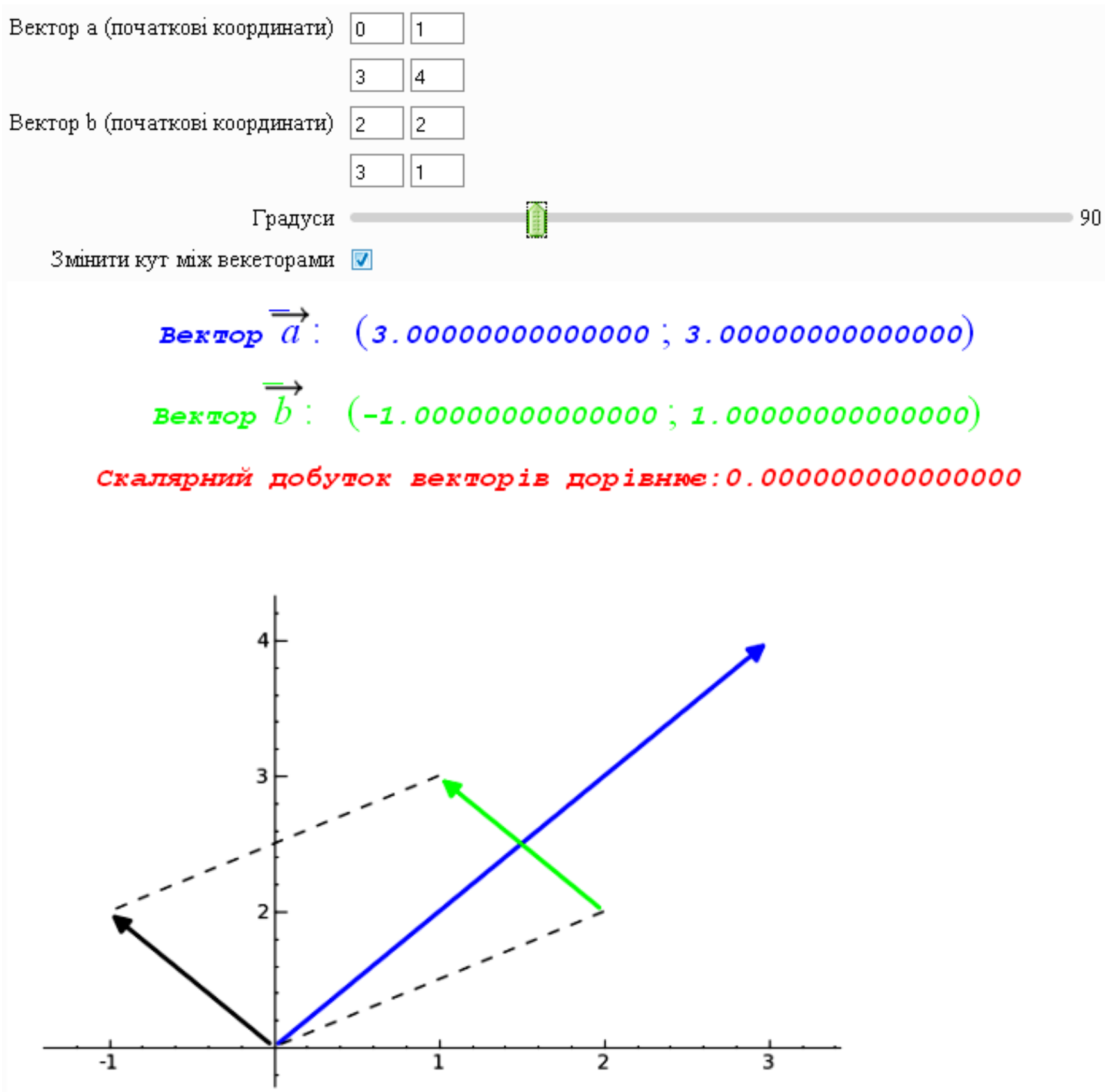

Рис. 3. Інтерфейс користувача моделі «Скалярний добуток векторів»

Наступну модель, пропонуємо використати при вивченні модуля «Ряди», зокрема розкладання елементарних функцій у ряд Маклорена.

При роботі з цією моделлю користувач має можливість змінювати функцію, що розкладається та вибирати, яку кількість частинних сум зображати на графіку. Запропонована модель може виступати не тільки в якості ілюстрації теоретичних понять, а й інструментом для дослі-джень. Так, змінюючи положення повзунка в ту чи іншу сторону, сту-дент помічає певну закономірність та робить висновок, що чим більшу кількість членів ряду Маклорена взяти, тим точніше графік відповідної частинної суми співпадає з графіком заданої функції.

Крім цього, студентам пропонується відстежити та пояснити насту-пний факт: якщо поступово (крок за кроком) збільшувати кількість час-тинних сум, користувач легко зрозуміє, що коли розкладати непарну функцію, то графік будь-якої парної частинної суми співпадає з передуючим йому графіком непарної частинної суми та навпаки.

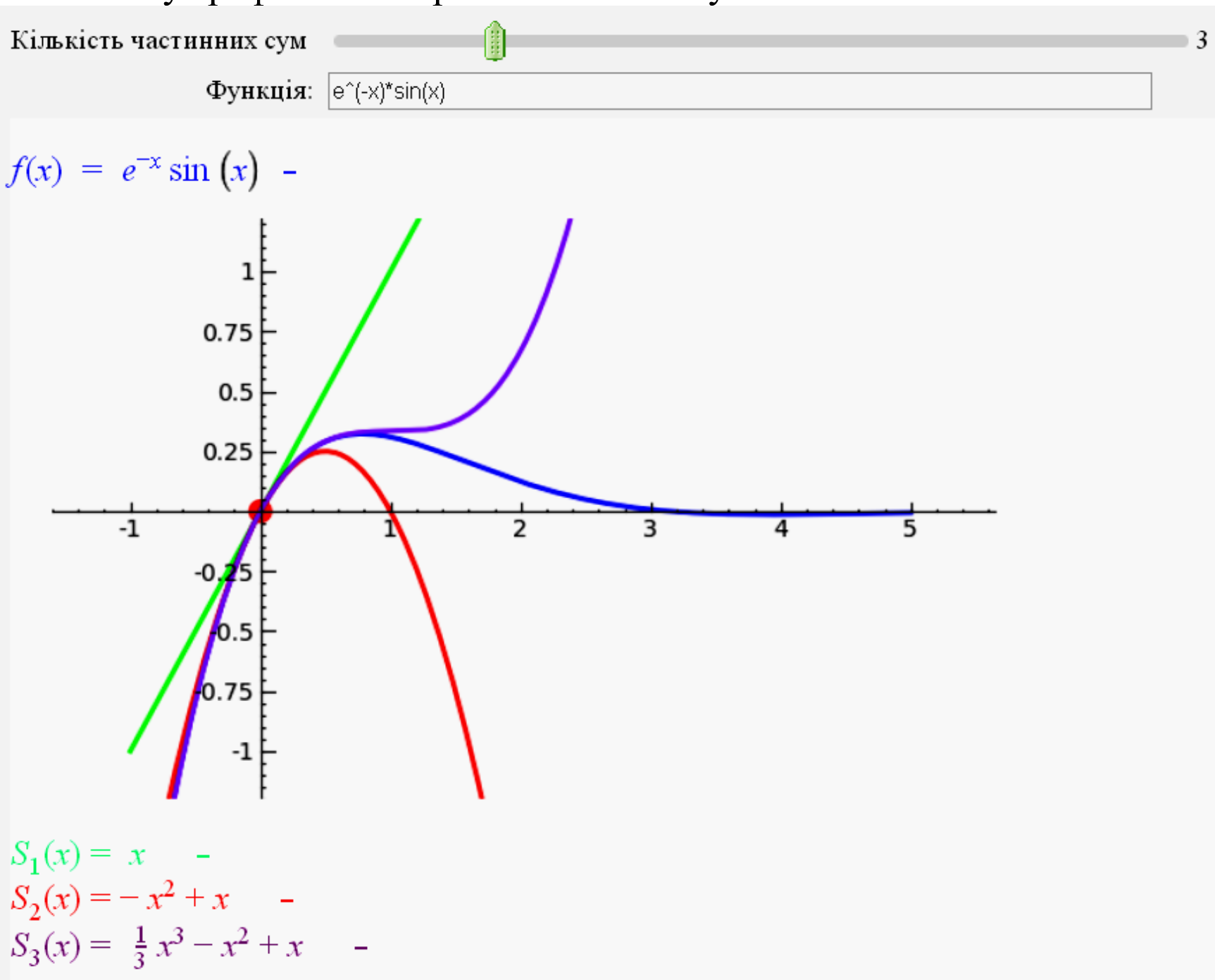

Рис. 4. Інтерфейс користувача моделі «Розвинення функції ряд Маклорена»

Таким чином, під час вивчення мобільного курсу вищої математики використання розроблених моделей дозволяє підвищити ефективність вивчення багатьох математичних понять; сприяє активізації пізнавальної активності та дослідницької діяльності студентів. Разом з тим, потрібно усвідомлювати, що ефективність впровадження у навчання СКМ забезпечується педагогічно виваженим добором змісту, методів і засобів навчання, зокрема комп'ютерних програм, форм і методів їх використання та систематичністю роботи студентів із комп'ютером [3].

Література